\documentclass[english,10pt, aps, prd, a4paper, preprintnumbers,floatfix, nofootinbib, showpacs, superscriptaddress, notitlepage]{revtex4-1} 
 \pdfoutput=1

\usepackage{graphicx}
\usepackage{rotate}

\usepackage[utf8]{inputenc} 

\usepackage{amsmath}
\usepackage{amssymb}
\usepackage{float}
\usepackage{comment}

\usepackage{soul}

\usepackage[usenames,dvipsnames]{color}
\usepackage[colorinlistoftodos]{todonotes}

 \usepackage{array}

\usepackage{amsmath}
\usepackage{amssymb}
\usepackage[colorlinks=true,citecolor=darkred,urlcolor=darkred, pdfborder={0 0 0}]{hyperref}
\usepackage[normalem]{ulem}
\newcommand {\ignore}[1]{}

\definecolor{darkred}{rgb}{0.6,0,0}

\def\tt1{$\mathrm{SU(3) \otimes SU(3)_L \otimes U(1)}$ }
\def\3311{$\mathrm{SU(3) \otimes SU(3)_L \otimes U(1)_X \otimes U(1)_{N}}$ }
\def\0331{$\mathrm{SU(3) \otimes SU(3)_L \otimes U(1)_X }$ }
\def\gsim{\raise0.3ex\hbox{$\;>$\kern-0.75em\raise-1.1ex\hbox{$\sim\;$}}}
\def\lsim{\raise0.3ex\hbox{$\;<$\kern-0.75em\raise-1.1ex\hbox{$\sim\;$}}}

\newcommand{\sm}{{Standard Model }}

\definecolor{mightnightblue}{RGB}{25,25,112}
\definecolor{brown}{rgb}{0.59, 0.29, 0.0}

\def\vev#1{\left\langle #1\right\rangle}

\def\21{$\mathrm{SU(2)_L \otimes U(1)_Y}$}

\def\sm{standard model }

\newcommand {\black} {\color{black}}

\newcommand{\AddrAHEP}{%
  AHEP Group, Institut de F\'{i}sica Corpuscular --
  CSIC/Universitat de Val\`{e}ncia, Parc Cient\'ific de Paterna.\\
 C/ Catedr\'atico Jos\'e Beltr\'an, 2 E-46980 Paterna (Valencia) - SPAIN}

\begin{document}

\bibliographystyle{unsrt}   

\title{Scotogenic dark matter stability from  gauged matter parity}
\author{Sin Kyu Kang} \email{skkang@seoultech.ac.kr}
\affiliation{School of Liberal Arts, Seoul-Tech, Seoul 139-743, Korea}
\affiliation{Institute of Convergence Fundamental Studies,  Seoul National University of Science and Technology, Seoul 139-743, Korea}
\author{Oleg Popov} \email{opopo001@ucr.edu}
\affiliation{Institute of Convergence Fundamental Studies,  Seoul National University of Science and Technology, Seoul 139-743, Korea}
\author{Rahul Srivastava}\email{rahulsri@ific.uv.es}
\affiliation{\AddrAHEP}
\author{Jos\'{e} W. F. Valle}\email{jose.valle@ific.uv.es}
\affiliation{\AddrAHEP}
\author{Carlos A. Vaquera-Araujo}\email{vaquera@fisica.ugto.mx}
\affiliation{Departamento de F\'isica, DCI, Campus Le\'on, Universidad de
Guanajuato, Loma del Bosque 103, Lomas del Campestre C.P. 37150, Le\'on, Guanajuato, M\'exico}
\affiliation{Consejo Nacional de Ciencia y Tecnolog\'ia, Av. Insurgentes Sur 1582. Colonia Cr\'edito Constructor, Del. Benito Ju\'arez, C.P. 03940, Ciudad de M\'exico, M\'exico}
\date{\today }

\begin{abstract}
\vspace{1cm} 

We explore the idea that dark matter stability results from the presence of a matter-parity symmetry,  arising naturally as a consequence of the spontaneous breaking of an extended $\mathrm{SU(3) \otimes SU(3)_L \otimes U(1)_X \otimes U(1)_{N}}$ electroweak gauge symmetry with fully gauged B-L. 
Using this framework we construct a theory for scotogenic dark matter and analyze its main features.

\end{abstract}

\maketitle

\section{Introduction}
\label{sec:intro}

Unveiling the nature of dark matter constitutes a big challenge in astroparticle physics, requiring the existence of new particles and also suggesting the presence of new symmetries capable of stabilizing the corresponding candidate particle on cosmological scales. 
A popular class of dark matter candidates in agreement with astrophysical and cosmological observations are the so-called Weakly Interacting Massive Particles, or WIMPs. For example, they are realized within supersymmetric extensions of the standard model~\cite{Jungman:1995df}. In that case stability follows from a postulated $Z_2$ symmetry called R-parity which also avoids fast proton decay and neutrino masses.

WIMPS however, arise in many other ways including ``low-scale'' models of neutrino mass generation~\cite{Boucenna:2014zba}, such as scotogenic dark matter~\cite{Ma:2006km} scenarios in which the exchange of ``dark sector particles'' is responsible for the radiative origin of neutrino mass. In such attractive scenarios WIMP dark matter emerges as radiative neutrino mass messenger~\cite{Hirsch:2013ola,Merle:2016scw}. 
In Refs.~\cite{Alves:2016fqe,Dong:2017zxo} it was suggested that an extended gauge symmetry can provide a natural setting for a theory of cosmological dark matter. 
The associated electroweak extensions both involve the SU(3)$_{\rm L}$ symmetry which has a long history. It is well-motivated due to its ability to ``explain'' the number of families to match that of colors, as a result of the anomaly cancellation requirement~\cite{Singer:1980sw,Pisano:1991ee,Frampton:1992wt}. For recent papers see Refs.~\cite{Hernandez:2013mcf,Hernandez:2014lpa,CarcamoHernandez:2017cwi,Barreto:2017xix}. These theories can also be made consistent with unification~\cite{Deppisch:2016jzl} and/or with the understanding of parity as a spontaneously broken symmetry~\cite{Hati:2017aez}. 
The two different models in~\cite{Alves:2016fqe,Dong:2017zxo} employ an extended electroweak gauge symmetry and the dark matter stability results from the presence of a matter-parity symmetry, $M_P$, a non-supersymmetric version of R-parity, that arises naturally as a consequence of the spontaneous breaking of the extended gauge symmetry.

The purpose of this letter is to go a step further along this idea. For definiteness we set out to explore the \3311 model proposed in ~\cite{Alves:2016fqe}
as a possible template for a theory of scotogenic dark matter. 
To do so we consider an extension of the original model containing extra vector-like fermions as well as scalars. These naturally contain the new messenger dark sector particles required to implement the scotogenic scenario. In Sect.~\ref{sec:model} we setup the stage for the theory, discussing the important issue of anomaly cancellation (details presented in appendix~\ref{sec:anomaly-cancellation}). In Secs~\ref{sec:neutrino-masses} and \ref{sec:masses} we study the loop-induced neutrino masses as well as the scalar boson and fermion mass spectra. In Sect~\ref{sec:dm-pheno} we briefly discuss the dark matter phenomenology, and conclude in the last section.

\section{Model}
\label{sec:model}

We consider a variant of the model introduced in \cite{Alves:2016fqe}, 
based on the \3311 gauge symmetry (3-3-1-1 for short). This is an abelian extension of the class of models based on the \tt1 gauge symmetry and as such, it inherits many of the defining features of these models. The  main motivation for the inclusion of the extra U(1)$_N$ symmetry is to allow for a fully gauged $B-L$ symmetry within a 3-3-1 framework \cite{Dong:2014wsa,Dong:2015yra}. In the present model, electric charge and $B-L$ are embedded into the gauge symmetry as 
\begin{align}
    Q & = T_3-\frac{T_8}{\sqrt{3}}+X,\\
    B-L & = -\frac{2}{\sqrt{3}}T_8+N,
\end{align}
with $T_i$ $(i = 1,2,3,...,8)$, $X$ and $N$ as the respective generators of SU(3)$_L$, U(1)$_X$ and U(1)$_N$. 

Under suitable conditions, the spontaneous symmetry breaking (SSB) pattern is such that a residual discrete symmetry arises from the $B-L$ symmetry breakdown. The role of the remnant symmetry is analogous to that of  $R$-parity in supersymmetric theories, we call it matter parity, $M_P=(-1)^{3(B-L)+2s}$. It follows that the stability of the lightest $M_P$-odd particle leads to a potentially viable WIMP dark matter candidate. For recent related papers see Ref.~\cite{Bonilla:2018ynb,CentellesChulia:2019gic,Ma:2019yfo}.

Here we show how the natural $M_P$ symmetry described by the 3-3-1-1 models can be responsible for both the neutrino mass generation as well as for the stability of dark matter within a scotogenic scenario, without the need to impose any additional symmetry by hand \cite{Ma:2015xla}.

The particle content of the model is shown in Table \ref{tab:tab3311}.  Anomaly cancellation requires that, if left-handed leptons $l_{aL}$; $a =1,2,3$ transform as triplets under $SU(3)_L$, i.e.
\begin{equation}
l_{aL}=\begin{pmatrix}\nu_a\\e_a\\N_a\end{pmatrix}_L,
\end{equation}
then two generations of quarks $ q_{i L}$; $i = 1,2$ must transform as anti-triplets and one as a triplet~\cite{Singer:1980sw} 
\begin{equation}
 q_{i L}=\begin{pmatrix}d_i\\-u_i\\D_i\end{pmatrix}_L\qquad q_{3L}=\begin{pmatrix}u_3\\d_3\\U_3\end{pmatrix}_L,
\end{equation}
This choice ``explains'' the number of generations as three (the same as the number of colors), an interesting feature of this class of models. The quark sector interactions are the same as in the original model~\cite{Singer:1980sw}. 
\begin{table}[ht]
    \centering
    \caption{3311 model particle content ($a=1,2,3$ and  $i =1,2$ represent generation indices). 
 Note the non-standard charges of ``right handed neutrinos'' $\nu_R$.}
    \label{tab:tab3311}
    \begin{tabular}{!{\vrule width 1.1pt}c|c|c|c|c|c|c!{\vrule width 1.1pt}}
        \hline
\hspace{0.2cm} Field \hspace{0.2cm}& \hspace{0.2cm}SU(3)$_c$ \hspace{0.2cm}&\hspace{0.2cm} SU(3)$_L$ \hspace{0.2cm}& \hspace{0.2cm}U(1)$_X$ \hspace{0.2cm}& \hspace{0.2cm}U(1)$_N$  \hspace{0.2cm}&\hspace{0.2cm} $Q$\hspace{0.2cm} &\hspace{0.2cm} $M_P = (-1)^{3(B-L)+2s}$ \hspace{0.2cm}\\
\hline\hline
        $q_{i L}$ & {\bf 3} &$ \overline{{\mathbf 3}} $& 0 & 0 & $(-\frac{1}{3},\frac{2}{3},-\frac{1}{3})^T$&$(++-)^T$ \\
       $ q_{3L}$ & {\bf 3} & {\bf 3} & $\frac{1}{3}$ & $\frac{2}{3}$ & $(\frac{2}{3},-\frac{1}{3},\frac{2}{3})^T$&$(++-)^T$ \\
        $u_{aR}$ & {\bf 3} & {\bf 1} & $\frac{2}{3}$ & $\frac{1}{3}$ &  $\frac{2}{3}$&$+$ \\
        $d_{aR}$ & {\bf 3} & {\bf 1} & $-\frac{1}{3}$ & $\frac{1}{3}$ &  $-\frac{1}{3}$&$+$ \\
        $U_{3R}$ & {\bf 3} & {\bf 1} & $\frac{2}{3}$ & $\frac{4}{3}$ &$\frac{2}{3}$&$-$ \\
        $D_{i R}$ & {\bf 3} & {\bf 1} & $-\frac{1}{3}$ & $-\frac{2}{3}$ & $-\frac{1}{3}$&$-$ \\
        $l_{aL}$ & {\bf 1} & {\bf 3} & $-\frac{1}{3}$ & $-\frac{2}{3}$ &$(0,-1,0)^T$&$(++-)^T$ \\
       $ e_{a R}$ & {\bf 1} & {\bf 1} & $-1$ & $-1$ & $-1$&$+$ \\
       \hline
       $ \nu_{i R}$ & {\bf 1} & {\bf 1} & $0$ & $-4$ & $0$&$-$ \\
       $  \nu_{3R}$ & {\bf 1} & {\bf 1} & $0$ & $5$ & $0$&$+$ \\
       $F_{a L,R}$ & {\bf 1} & {\bf 3} & $-\frac{1}{3}$ & $-\frac{1}{3}$ & ($0,-1,0$)&($--+$) \\
       \hline\hline
        $\eta$ & {\bf 1} & {\bf 3} & $-\frac{1}{3}$ & $\frac{1}{3}$ &  $(0,-1,0)^T$&$(++-)^T$ \\
        $\rho$ & {\bf 1} & {\bf 3} & $\frac{2}{3}$ & $\frac{1}{3}$ &  $(1,0,1)^T$&$(++-)^T$ \\
        $\chi$ & {\bf 1} & {\bf 3} & $-\frac{1}{3}$ & $-\frac{2}{3}$ &  $(0,-1,0)^T$&$(--+)^T$ \\
        $\phi$ & {\bf 1} & {\bf 1} & $0$ & $2$ & $0$&$+$ \\
        \hline
        $S$ & {\bf 1} & {\bf 1} & $0$ & $\frac{2}{3}$ &  $0$&$+$ \\
        $\sigma$ & {\bf 1} & {\bf 1} & $0$ & $\frac{1}{3}$ &  $0$&$-$ \\
        $\Omega$ & {\bf 1} & {\bf 6} & $\frac{2}{3}$ & $\frac{2}{3}$ & $\left(\begin{matrix}
    0 & 1 & 0 \\
    1 & 2 & 1 \\
    0 & 1 & 0\end{matrix}\right)$&$ \left(\begin{matrix}
    + & + & - \\
    + & + & - \\
    - & - & + \end{matrix}\right)$\\
\hline    
\end{tabular}
\end{table}

The new ingredients of the model, with respect to \cite{Alves:2016fqe}, are the vector-like fermion triplets $F_{a L,R}$~\footnote{These are called vector-like insofar as their gauge charges are concerned, required for anomaly cancellation. However, as we will see later, they have Majorana as well as Dirac-type mass terms.}, and the extended scalar sector spanned by $S$, $\sigma$ and $\Omega$. These fields will be responsible for the neutrino mass generation mechanism described in the next section. The Yukawa terms involving leptons and vector-like fermions are thus given by
\begin{align}
\mathcal{L}^{\text{Yuk}}&\supset\overline{l}_{aL}^i Y_{e}^{ab}e_{bR}\rho_i+\overline{F}_{aR}^i Y_{1}^{ab}l_{ibL}\sigma +\overline{F}_{aL} m^{ab}_F F_{bR}+ F_{iaL,R}Y_{2L,R}^{ab}F_{jbL,R}\Omega^{ij}+\text{H.c.},
\end{align}
where $i,j=1,2,3$ are $SU(3)_L$ indices. 

Notice the unconventional chiral charges of $\nu_R$ fields, owing to which the tree level coupling between $l_{aL}$ and  $\nu_R$ is automatically avoided. 
Such chiral solutions were already known in context of $B-L$ symmetry~\cite{Montero:2007cd,Ma:2014qra,Ma:2015raa,Ma:2015mjd}. Here we show for the first time that they can also be embedded inside bigger gauge groups containing $B-L$ symmetry.
In appendix~\ref{sec:anomaly-cancellation} we display explicitly the non-trivial way in which the anomalies involving the U(1) gauge symmetries cancel, despite the unconventional $\nu_R$ charge assignments.

After the singlet scalar $\phi$ develops a vacuum expectation value (VEV), the gauged $B-L$ symmetry is spontaneously broken by two units, leaving a discrete remnant symmetry $M_P=(-1)^{3(B-L)+2s}$.  The most general VEV alignment for the scalar triplets and $\phi$, which is compatible with the preservation of $M_P$ symmetry, is given by
\begin{equation}
\langle\eta\rangle=\frac{1}{\sqrt{2}}(v_1,0,0)^T,\quad \langle\rho\rangle=\frac{1}{\sqrt{2}}(0,v_2,0)^T, \quad\langle\chi\rangle=(0,0,w)^T, \quad \langle\phi\rangle=\frac{1}{\sqrt{2}}\Lambda.
\end{equation}
Furthermore, if the VEV alignment for the scalars $S$, $\sigma$ and $\Omega$  is
\begin{equation}
\langle S\rangle= v_s, \quad\langle \sigma\rangle= 0, \quad \langle \Omega\rangle=
\left(\begin{matrix}
    w_1 & 0 & 0 \\
    0 & 0 & 0 \\
    0 & 0 & w_2\end{matrix}\right),
\end{equation}
then, $M_P$ is an exactly conserved symmetry.

Assuming $w, \Lambda, w_2, v_s \gg v_1, v_2,w_1 $ the spontaneous symmetry breaking (SSB) pattern of the model is given by
\begin{align}
SU(3)_{C}^{}\times &SU(3)_{L}^{}\times U(1)_{X}^{}\times U(1)_{N}^{} \nonumber\\
&\downarrow  w, \Lambda, w_2, v_s \nonumber\\
SU(3)_C^{} &\times SU(2)_{L}^{}\times U(1)_{Y}^{}\times M_P^{} \nonumber\\
&\downarrow v_1,v_2, w_1 \nonumber\\
 SU(3)_C^{} & \times U(1)_{Q}^{}\times M_P^{}\,,
\end{align} 
and the phenomenology for quarks, charged leptons and gauge bosons of the model coincides largely with the analysis performed in \cite{Dong:2014wsa}.

\section{Neutrino masses}
\label{sec:neutrino-masses}

First we notice that, thanks to the charges of the scalars in the model as well as the unusual assignments of $\nu_R$ charges, tree level neutrino masses are absent in this model.
These include a tree-level Dirac-like mass term coupling the electrically neutral isodoublet and isosinglet members of the lepton triplets~\cite{Valle:1983dk}.
Likewise, the absence of genuine right-handed neutrino fields implies no tree-level seesaw-type neutrino mass contributions, such as the type-I Majorana seesaw used in Ref.~\cite{Alves:2016fqe} or the type-II Dirac seesaw proposed in Ref.~\cite{Reig:2016ewy}.
Matter-parity conservation also forbids seesaw-type neutrino Majorana masses mediated by the vector-like fermions.

As a result small neutrino masses are generated only at the one-loop level, mediated by the vector-like fermions $F_{L,R}$, the singlet scalars $S$, $\sigma$ and the scalar sextet $\Omega$. The relevant interactions among these fields are
\begin{align}
    \mathcal{L}_{\text{\tiny{m$_{\nu}$}}}&=\overline{F}_{aR}^i Y_{1}^{ab}l_{ibL}\sigma + F_{iaR}Y_{2R}^{ab}F_{jbR}\Omega^{ij} +\mu_2\sigma^2 S^* + \text{H.c.}
    \label{eq:neut-lag}
\end{align}
where $i,j$ represent $SU(3)_L$ indices. 

\begin{figure}[h]
 \centering
\includegraphics[scale=1,trim={6cm 21cm 6cm 3cm},clip]{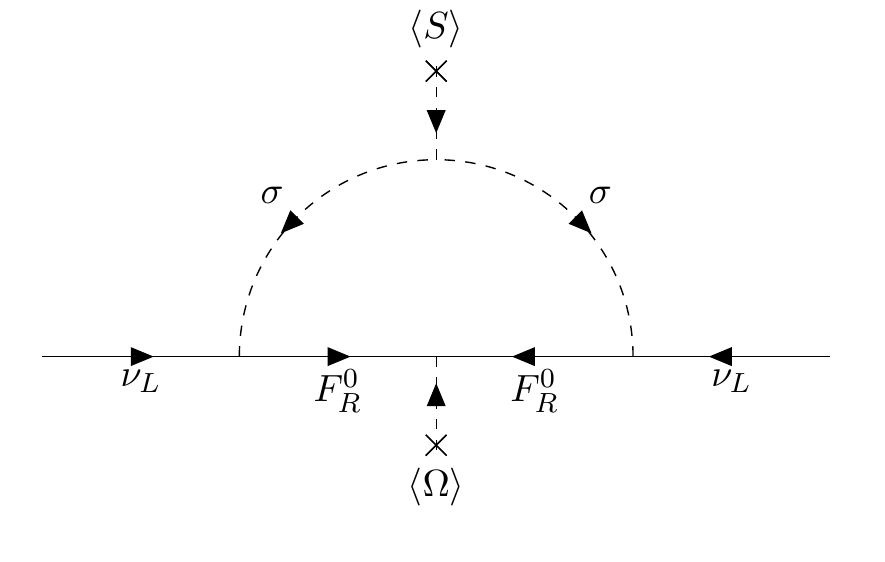}
 \caption{3311 model scotogenic neutrino mass.}
 \label{fig:mnu3311_1}
 \end{figure}

Figure \ref{fig:mnu3311_1}  depicts the one-loop diagram for light $\nu_L$ masses. In the neutrino mass diagram, the fields running in the loop ($F_{R}^{0},\sigma$)  have odd matter parity transformation, $M_P$, whereas the fields that appear outside the loop ($\nu_L,\Omega_{11},S$) are even under matter parity transformation.

The resulting neutrino radiative mass is given as 
\small{
\begin{align}
\label{eq:mnu}
    m_{\nu}^{ab}&=\frac{Y_{1}^{ac}}{32\pi^2} \left\{s_{F-}^2m_{N_1}\left[Z\left(\frac{m_{\xi_{1R}}^2}{m_{N_1}^2}\right)c_{\xi R}^2+Z\left(\frac{m_{\xi_{2R}}^2}{m_{N_1}^2}\right)s_{\xi R}^2-Z\left(\frac{m_{\xi_{1I}}^2}{m_{N_1}^2}\right)c_{\xi I}^2-Z\left(\frac{m_{\xi_{2I}}^2}{m_{N_1}^2}\right)s_{\xi I}^2\right]\right. \nonumber \\
    &\left.+c_{F-}^2m_{N_2}\left[Z\left(\frac{m_{\xi_{1R}}^2}{m_{N_2}^2}\right)c_{\xi R}^2+Z\left(\frac{m_{\xi_{2R}}^2}{m_{N_2}^2}\right)s_{\xi R}^2-Z\left(\frac{m_{\xi_{1I}}^2}{m_{N_2}^2}\right)c_{\xi I}^2-Z\left(\frac{m_{\xi_{2I}}^2}{m_{N_2}^2}\right)s_{\xi I}^2\right]\right\}_{cd} Y_1^{db},
\end{align}
}
where the loop fuction $Z(x)$ is defined as 
\begin{align}
Z(x)&=\frac{x}{1-x}\text{ln}x,
\end{align}
and mixing angles of $\left(\eta_3, \sigma\right)_{R,I}$ and the odd component of $\left(F_L^{0},F_R^{c0}\right)$ are obtained from Eqs.~\ref{eq:etasigma_theta} and~\ref{eq:f0m_theta}, respectively. 

Notice that the fields $\nu_R$ remain massless after spontaneous symmetry breaking, and do not play a direct role in neutrino mass generation. They can contribute as extra degrees of freedom in primordial Big Bang nucleosyhthesis. However, this is not an issue since consistency with cosmological observations in such a case can be ensured by having the extra gauge bosons adequately heavy~\cite{GonzalezGarcia:1989py}. Alternatively, these extra fermions could be made massive trivially through the inclusion of extra scalar singlets with appropriate U(1)$_N$ charges.

\section{Mass Spectrum}
\label{sec:masses}

The full scalar potential of the model is written as
\begin{align}
\label{eq:3311_1V}
    &V=V_o+\lambda_{\eta\rho 2}\left(\eta^{\dagger}\rho\right)\left(\rho^{\dagger}\eta\right)+\lambda_{\eta\chi 2}\left(\eta^{\dagger}\chi\right)\left(\chi^{\dagger}\eta\right)+\lambda_{\eta\Omega 2}\eta^{\dagger i}\Omega^{\dagger}_{ij}\Omega^{jk}\eta_{k}+\lambda_{\chi\rho 2}\left(\chi^{\dagger}\rho\right)\left(\rho^{\dagger}\chi\right)\nonumber \\
    &+\lambda_{\rho\Omega 2}\rho^{\dagger i}\Omega^{\dagger}_{ij}\Omega^{jk}\rho_{k}+\lambda_{\chi\Omega 2}\chi^{\dagger i}\Omega^{\dagger}_{ij}\Omega^{jk}\chi_{k}+\lambda_{\Omega 2}\Omega^{ij}\Omega^{kl}\Omega_{\alpha\beta}^{\dagger}\Omega_{\gamma\delta}^{\dagger}\epsilon_{ikm}\epsilon_{jln}\epsilon^{\alpha\gamma m}\epsilon^{\beta\delta n}\nonumber \\
    &+\mu_1\eta_i\rho_j\chi_k\epsilon^{ijk}+\mu_2\sigma^2 S^* +\lambda_1\chi_i\Omega^{ij}\chi_j S+\lambda_{2}\sigma^*\eta_i\Omega^{ij}\chi_{j}+\lambda_3\phi^* S^3+\lambda_4 S\sigma \left(\eta^{\dagger}\chi\right)+\text{H.c.}
\end{align}
where the $V_o$ piece consists of the following terms,
\begin{align}
V_o&=\sum_{\substack{x\in \left(\eta,\rho,\chi,\right.\\ \left.\phi,\Omega,\sigma,S\right)}} m_x^2\left(x^{\dagger}x\right)+\sum_{\substack{x\in \left(\eta,\rho,\chi,\right. \\ \left.\phi,\Omega,\sigma,S\right)}} \frac{\lambda_x}{2}\left(x^{\dagger}x\right)^2+\sum_{\substack{x,y\in \left(\eta,\rho,\chi,\right. \\ \left.\phi,\Omega,\sigma,S\right)\land x>y}} \lambda_{xy}\left(x^{\dagger}x\right)\left(y^{\dagger}y\right).
\end{align}

The conditions for the minimization  of the scalar potential are given as follows:
\begin{align}
\left.\frac{\partial V}{\partial \eta_1}\right|_{\eta_1\rightarrow v_1}&=0\implies 2 m^2_{\eta}+\lambda_{\eta\Omega 2}w_1^2+\lambda_{\eta\Omega}\left(w_1^2+w_2^2\right)+\lambda_{\eta s}v_s^2+\lambda_{\eta}v_1^2+\lambda_{\eta\rho}v_2^2+\lambda_{\eta\phi}\Lambda^2+\lambda_{\eta\chi}w^2+\sqrt{2}\mu_1 \frac{v_2 w}{v_1}=0\\
\left.\frac{\partial V}{\partial \rho_2}\right|_{\rho_2\rightarrow v_2}&=0\implies 2 m^2_{\rho}+\lambda_{\rho\Omega}\left(w_1^2+w_2^2\right)+\lambda_{\rho s}v_s^2+\lambda_{\eta\rho}v_1^2+\lambda_{\rho}v_2^2+\lambda_{\rho\phi}\Lambda^2+\lambda_{\rho\chi}w^2+\sqrt{2}\mu_1 \frac{v_1 w}{v_2}=0\\
\left.\frac{\partial V}{\partial \chi_3}\right|_{\chi_3\rightarrow w}&=0\implies 2 m^2_{\chi}+\lambda_{\chi\Omega}\left(w_1^2+w_2^2\right)+\lambda_{\chi s}v_s^2+\lambda_{\rho\chi}v_2^2+\lambda_{\chi}w^2+\lambda_{\chi\phi}\Lambda^2+\lambda_{\chi\eta}v_1^2+\lambda_{\chi\Omega 2}w_2^2+ 2 \lambda_1w_2 v_s +\sqrt{2}\mu_1 \frac{v_1 v_2}{w}=0\\
\left.\frac{\partial V}{\partial \phi}\right|_{\phi\rightarrow \Lambda}&=0\implies 2 m^2_{\phi}+\lambda_{\phi}\Lambda^2+\lambda_{\phi s}v_s^2+\lambda_{\phi\eta}v_1^2+\lambda_{\phi\rho}v_2^2+\lambda_{\phi\chi}w^2+\lambda_{\phi\Omega}\left(w_1^2+w_2^2\right)+\lambda_3 \frac{v_s^3}{\Lambda}=0\\
\left.\frac{\partial V}{\partial \Omega_{11}}\right|_{\Omega_{11}\rightarrow w_1}&=0\implies 2 m^2_{\Omega}+\lambda_{\Omega 2}w_2^2+\lambda_{\Omega}\left(w_1^2+w_2^2\right)+\lambda_{\Omega s}v_s^2+\left(\lambda_{\Omega\eta}+\lambda_{\Omega\eta 2}\right)v_1^2+\lambda_{\Omega\rho}v_2^2+\lambda_{\Omega\chi}w^2+\lambda_{\Omega\phi}\Lambda^2=0\\
\left.\frac{\partial V}{\partial \Omega_{33}}\right|_{\Omega_{33}\rightarrow w_2}&=0\implies \lambda_{\Omega 2}\left(w_1^2-w_2^2\right)-\lambda_{\Omega\eta 2}v_1^2+\lambda_{\Omega\chi 2}w^2+\lambda_1 \frac{v_s w^2}{w_2}=0\\
\left.\frac{\partial V}{\partial S}\right|_{S\rightarrow v_s}&=0\implies 2 m_s^2+\lambda_{s\Omega}\left(w_1^2+w_2^2\right)+\lambda_{s}v_s^2+\lambda_{s\eta}v_1^2+\lambda_{s\rho}v_2^2+\lambda_{s\phi}\Lambda^2+\lambda_{s\chi}w^2+3\lambda_3 v_s \Lambda+\lambda_1 \frac{w_2 w^2}{v_s}=0
\end{align}

Notice that, due to the assumed positivity of its squared mass, the field $\sigma$ has zero vacuum expectation value, as required for the conservation of the matter parity symmetry. \\

\subsection{Scalar masses}
\label{sec:scalar-masses}

The physical scalars include the following particles, classified according to their electric charges and matter parities:
\begin{itemize}
\item $Q=\pm 2, M_P=+$: consists of only $\Omega_{22}^{\pm 2}$ complex scalar, its mass is given by
\begin{equation}
\label{eq:m22p}
m_{\scriptstyle{\Omega_{22}}}^2=\frac{1}{2}\left(\lambda_{\eta\Omega 2}w_2^2v_1^2+\lambda_{\rho\Omega 2}\left(w_1^2-w_2^2\right)v_2^2-w_1^2 w^2\left(\lambda_{\chi\Omega 2}+\lambda_1 \frac{v_s}{w_2}\right)\right)\left(w_1^2-w_2^2\right)^{-1}.
\end{equation}\\
\item $Q=\pm 1, M_P=+$: consists of two complex physical scalar eigenstates and one complex Nambu-Goldstone (NG) boson corresponding to I-spin of $SU(3)_L$ gauge group (charged boson connecting $(T_3,T_8)$ states $(\frac{1}{2},\frac{1}{2\sqrt{3}})\leftrightarrow(-\frac{1}{2},\frac{1}{2\sqrt{3}})$). Corresponding mass squared matrix is given by
\begin{align}
\label{eq:m21p}
\frac{1}{2}\left(\begin{matrix}
\lambda_{\eta\rho 2}v_2^2-\lambda_{\eta\Omega 2}w_1^2-\sqrt{2}\mu_1\frac{v_2 w}{v_1} & \lambda_{\eta\rho 2}v_1 v_2-\sqrt{2}\mu_1 w & \lambda_{\eta\Omega 2}w_1 v_1 \\
\lambda_{\eta\rho 2}v_1 v_2-\sqrt{2}\mu_1 w & \lambda_{\rho\Omega 2}w_1^2+\lambda_{\eta\rho 2}v_1^2-\sqrt{2}\mu_1\frac{v_1 w}{v_2} & \lambda_{\rho\Omega 2} w_1 v_2 \\
\lambda_{\eta\Omega 2}w_1 v_1 & \lambda_{\rho\Omega 2} w_1 v_2 & \lambda_{\rho\Omega 2} v_2^2-\lambda_{\eta\Omega 2}v_1^2
\end{matrix}\right)
\end{align}
in the basis $(\eta_2^c,\rho_1,\Omega_{12})$.\\
\item $Q=\pm 1, M_P=-$: consists of two complex physical scalar eigenstates and one complex Nambu-Goldstone (NG) boson corresponding to V-spin of $SU(3)_L$ gauge group (charged boson connecting $(T_3,T_8)$ states $(-\frac{1}{2},\frac{1}{2\sqrt{3}})\leftrightarrow(0,-\frac{1}{\sqrt{3}})$). Corresponding mass squared matrix is given by
\begin{align}
\label{eq:m21m}
\frac{1}{2}\left(\begin{matrix}
\lambda_{\rho\Omega 2}w_2^2+\lambda_{\rho\chi 2}w^2 & \lambda_{\rho\chi 2}v_2 w-\sqrt{2}\mu_1 v_1 & \lambda_{\rho\Omega 2}w_2 v_2 \\
\lambda_{\rho\chi 2}v_2 w-\sqrt{2}\mu_1 v_1 & \lambda_{\rho\chi 2}v_2-\lambda_{\chi\Omega 2}w_2^2-2 \lambda_{1}w_2 v_s-\sqrt{2}\mu_1 \frac{v_1 v_2}{w} & \lambda_{\chi\Omega 2} v_2 w+2\lambda_1 v_s w \\
\lambda_{\rho\Omega 2}w_2 v_2 & \lambda_{\chi\Omega 2} v_2 w+2\lambda_1 v_s w & \lambda_{\rho\Omega 2}v_2-\lambda_{\chi\Omega 2}w^2 - 2\lambda_1 \frac{v_s w^2}{w_2}
\end{matrix}\right)
\end{align}
in the basis $(\rho_3,\chi_2^c,\Omega_{23})$.\\
\item $Q=0, M_P=+$: The CP even part consists of 7 physical scalar eigenstates. Corresponding mass squared matrix is given in Eq.~\ref{eq:m20pR} of Ap.~\ref{sec:mass_det}. CP odd part consists of 4 pseudo-scalars and 3 NG bosons. Corresponding mass squared matrix is given in Eq.~\ref{eq:m20pI} of Ap.~\ref{sec:mass_det}.\\
\item $Q=\pm 0, M_P=-$: consists of three real physical scalar and pseudo-scalar eigenstates and one complex Nambu-Goldstone (NG) boson corresponding to U-spin of $SU(3)_L$ gauge group (charged boson connecting $(T_3,T_8)$ states $(\frac{1}{2},\frac{1}{2\sqrt{3}})\leftrightarrow(0,-\frac{1}{\sqrt{3}})$). Corresponding CP-even mass squared matrix is given in Eq.~\ref{eq:m20mR} of Ap.~\ref{sec:mass_det}. The corresponding CP-odd mass squared matrix is given in Eq.~\ref{eq:m20mI} of Ap.~\ref{sec:mass_det}.
\end{itemize}
In order to calculate the neutrino masses we use the simplification $w, \Lambda, w_2, v_s \gg v_1, v_2,w_1 $, then the mass squared matrices in Eqs.~\ref{eq:m20mR} and~\ref{eq:m20mI} become block diagonalized 2-by-2 matrices and are given by
\begin{align}
\label{eq:m20mRS}
\frac{1}{2}&\left(\begin{matrix}
\lambda_{\eta\Omega 2}w_2^2+\left(\lambda_{\eta\chi 2}w-\sqrt{2}\mu_1\right)w & \left(\lambda_2 w_2+\lambda_4 v_s\right)w \\
\left(\lambda_2 w_2+\lambda_4 v_s\right)w & 2 m_{\sigma}^2+\lambda_{\sigma\Omega}w_2^2+\lambda_{s\sigma}v_s^2+\lambda_{\phi\sigma}\Lambda^2+\lambda_{\chi\sigma}w^2+2\sqrt{2}\mu_2 v_s
\end{matrix}\right)
\oplus \nonumber \\
&\left(\begin{matrix}
-\left(\lambda_{\chi\Omega 2}w_2+2\lambda_1 v_s\right)w_2 & \left(\lambda_{\chi\Omega 2}w_2+2\lambda_1 v_s\right)w \\
\left(\lambda_{\chi\Omega 2}w_2+2\lambda_1 v_s\right)w & -\lambda_{\chi\Omega 2}w^2-2\lambda_1 \frac{v_s w^2}{w_2}
\end{matrix}\right),
\end{align}
and 
\begin{align}
\label{eq:m20mIS}
\frac{1}{2}&\left(\begin{matrix}
\lambda_{\eta\Omega 2} w^2+\lambda_{\eta\chi 2} w^2-\sqrt{2}\mu_1 w & \left(\lambda_2 w_2 + \lambda_4 v_s\right)w \\
\left(\lambda_2 w_2 + \lambda_4 v_s\right)w & 2 m_{\sigma}^2+\lambda_{\sigma\Omega}w_2^2+\lambda_{s\sigma}v_s^2+\lambda_{\phi\sigma}\Lambda^2+\lambda_{\chi\sigma}w^2-2\sqrt{2}\mu_2 v_S
\end{matrix}\right)
\oplus \nonumber \\
&\left(\begin{matrix}
-w_2\left(\lambda_{\chi\Omega 2}w_2+2\lambda_1 v_s\right) & -w\left(\lambda_{\chi\Omega 2}w_2+2\lambda_1 v_s\right) \\
-w\left(\lambda_{\chi\Omega 2}w_2+2\lambda_1 v_s\right) & -\left(\lambda_{\chi\Omega 2}w_2+2\lambda_1 v_s\right)\frac{w^2}{w_2}
\end{matrix}\right),
\end{align}
respectively. The part relevant for neutrino masses is the first 2-by-2 block of mass squared matrices~\ref{eq:m20mRS} and~\ref{eq:m20mIS} in the basis $(\eta_{3R,I}, \sigma_{R,I})$. The eigenvalues are given by

\begin{align}
m_{\xi_{1,2R}}^2&=\frac{1}{4}\left[\left(2 m_{\sigma}^2+\lambda_{\eta\Omega 2}w_2^2+\lambda_{\eta\chi 2}w^2-2\sqrt{2}\mu_1 w +\lambda_{\sigma\Omega}w_2^2+\lambda_{s\sigma}v_s^2+\lambda_{\phi\sigma}\Lambda^2+\lambda_{\chi\sigma}w^2 + 2\sqrt{2}\mu_2 v_s\right)\right. \nonumber \\
&\pm\left.\sqrt{\left(-2 m_{\sigma}^2+\lambda_{\eta\Omega 2}w_2^2+\lambda_{\eta\chi 2}w^2-2\sqrt{2}\mu_1 w -\lambda_{\sigma\Omega}w_2^2-\lambda_{s\sigma}v_s^2-\lambda_{\phi\sigma}\Lambda^2-\lambda_{\chi\sigma}w^2 - 2\sqrt{2}\mu_2 v_s\right)^2+4\left(\lambda_2 w_2+\lambda_4 v_s\right)^2 w^2}\right],\\
m_{\xi_{1,2I}}^2&=\frac{1}{4}\left[\left(2 m_{\sigma}^2+\lambda_{\eta\Omega 2}w_2^2+\lambda_{\eta\chi 2}w^2-2\sqrt{2}\mu_1 w +\lambda_{\sigma\Omega}w_2^2+\lambda_{s\sigma}v_s^2+\lambda_{\phi\sigma}\Lambda^2+\lambda_{\chi\sigma}w^2 - 2\sqrt{2}\mu_2 v_s\right)\right. \nonumber \\
&\pm\left.\sqrt{\left(-2 m_{\sigma}^2+\lambda_{\eta\Omega 2}w_2^2+\lambda_{\eta\chi 2}w^2-2\sqrt{2}\mu_1 w -\lambda_{\sigma\Omega}w_2^2-\lambda_{s\sigma}v_s^2-\lambda_{\phi\sigma}\Lambda^2-\lambda_{\chi\sigma}w^2 + 2\sqrt{2}\mu_2 v_s\right)^2+4\left(\lambda_2 w_2+\lambda_4 v_s\right)^2 w^2}\right],
\end{align}

and mixing is given by
\begin{align}
\label{eq:etasigma_m}
\left(\begin{matrix}
\eta_{3} \\ \sigma
\end{matrix}\right)_{R,I}&=\left(\begin{matrix}
\text{cos}\theta & -\text{sin}\theta \\ \text{sin}\theta & \text{cos}\theta
\end{matrix}\right)_{R,I}\left(\begin{matrix}
\xi_1 \\ \xi_2
\end{matrix}\right)_{R,I},\\
\label{eq:etasigma_theta}
\text{tan}\theta_{R,I}&=\frac{\left(\lambda_2 w_2+\lambda_4 v_s\right) w}{\left(-2 m_{\sigma}^2+\lambda_{\eta\Omega 2}w_2^2+\lambda_{\eta\chi 2}w^2-2\sqrt{2}\mu_1 w -\lambda_{\sigma\Omega}w_2^2-\lambda_{s\sigma}v_s^2-\lambda_{\phi\sigma}\Lambda^2-\lambda_{\chi\sigma}w^2\mp 2\sqrt{2}\mu_2 v_s\right)}.
\end{align}

\vspace{1cm}

\subsection{Fermion masses}
\label{sec:fermion-masses}

First note that the fermions $N_{L,\alpha}=l_{3,\alpha}$ and $\nu_{\alpha,R}^{c}$ do not mix with others at the tree level. 
The fermions which are relevant for neutrino mass generation are the $M_P$ odd components of 
$(F_{L}^{0},F_R^{c,0})$, which lie in the sector with $(Q=0$, $M_P=-)$.
The corresponding mass matrix is given by
\begin{align}
\label{eq:fmm0m}
\left(\begin{matrix}
Y_{2L}w_1 & m_F^{\dagger} \\
m_F^{\dagger} & Y_{2R}^{\dagger} w_1
\end{matrix}\right)+\text{H.c.},
\end{align}
in the basis of $(F_{L}^{0},F_R^{c,0})$ of the $M_P$-odd components. Corresponding eigenvalues and mass eigenstates are given by
\begin{align}
\label{eq:fm0m}
m_{N_{1,2}}&=\frac{w_1}{2}\left[\left(Y_{2L}+Y_{2R}^{\dagger}\right)\pm \sqrt{\left(Y_{2L}-Y_{2R}^{\dagger}\right)^2+4 \left(\frac{m_F^{\dagger}}{w_1}\right)^2}\right],\\
\label{eq:f0m_mix}
\left(\begin{matrix}
F_{L}^{0} \\ F_{R}^{c0}
\end{matrix}\right)
&=\left(\begin{matrix}
\text{cos}\theta & -\text{sin}\theta \\ \text{sin}\theta & \text{cos}\theta
\end{matrix}\right)_f\left(\begin{matrix}
F_{1L} \\ F_{2R}^c
\end{matrix}\right),\\
\end{align}
and $\theta_f$ is given by
\begin{align}
\label{eq:f0m_theta}
\text{tan}\theta_f&=\frac{2m_F^{\dagger}}{w_1\left(Y_{2L}-Y_{2R}^{\dagger}\right)}.
\end{align}

\section{Dark Matter Phenomenology}
\label{sec:dm-pheno}


Before concluding, let us briefly discuss the phenomenology of dark matter in our model. As can be seen from the discussion of previous sections,  in our model dark matter is the mediator of neutrino mass generation. 
First notice that the stability of dark matter follows from the matter parity symmetry $M_P$, which is a residual symmetry of the full \3311 gauge symmetry. 
All the particles odd under matter parity $M_P$ belong to the ``dark sector'',  with the lightest amongst them being the dark matter candidate.
As can be seen from Table \ref{tab:tab3311} our model can have both fermionic or scalar dark matter, depending on which one is the lightest
\footnote{The gauge boson $X^0$ is also odd under $M_P$, hence a potential dark matter candidate. However, as discussed in \cite{Dong:2013wca}, 
it cannot be a viable one, since its relic density turns out to be too small. It follows that, in our model, $X^0$ cannot be the lightest dark sector particle.}.
As an illustrative example, here we briefly discuss the phenomenological constraints for the case of scalar dark matter.

 \begin{figure}[h!]
 \centering
\includegraphics[width=0.7\textwidth]{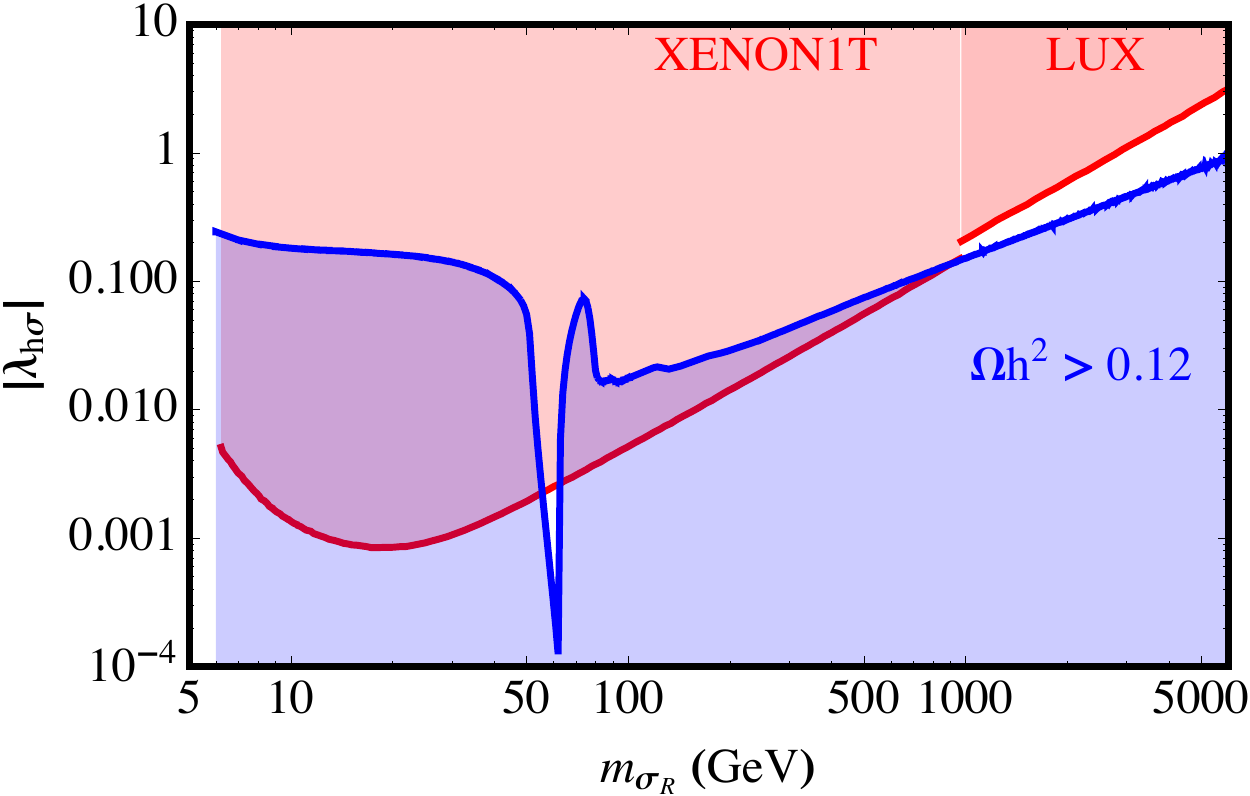}
\caption{The direct detection and relic abundance constraints on the dark matter mass $m_{\sigma_R}$ and its coupling $\lambda_{\sigma h}$ 
to the Higgs boson. The red shaded region is ruled out by direct detection experiments, XENON1T~\cite{Aprile:2018dbl} and LUX~\cite{Akerib:2016vxi}, while the blue shaded region is not compatible with the dark matter relic abundance~\cite{Aghanim:2018eyx}.  The combination of relic density and direct detection constraints implies that, apart from a tiny region near half Higgs mass, 
the mass of dark matter must lie in the TeV range. The plot is obtained for a specific benchmark, see text for details. }
 \label{fig:dm}
 \end{figure}

As can be seen from Table \ref{tab:tab3311} and Figure \ref{fig:mnu3311_1}, the $M_P$ odd scalar $\sigma$ takes part in the 
neutrino mass generation loop. Assuming that it is the dark matter particle we now analyze the associated phenomenology.
For simplicity we assume that the mixing between $\sigma$ and the other $M_P$ odd scalars is negligible.
In order for $\sigma$ to be dark matter it should also be the lightest particle amongst all dark sector particles.
Due to its $U(1)_N$ charge,  $\sigma$ must be a complex field with real ($\sigma_R$) and imaginary ($\sigma_I$) components. 
Owing to the $\mu_2$ coupling of Eq. \ref{eq:neut-lag}, the two masses cannot be exactly degenerate once the $S$ field get a vev.  
The vev of $S$ field breaks $U(1)_{B-L} \to \mathcal{Z}_2$ subgroup, the symmetry protecting the small neutrino mass, see Fig.~\ref{fig:mnu3311_1}.
%
In the limit of $\vev{S} \to 0$ the symmetry of the theory gets enhanced. 
%
The presence of $\mu_2$ term coupled with the $\vev{S}$ implies that $\sigma_R$ and   $\sigma_I$ components cannot be degenerate in mass and the lighter of the two will be the dark matter candidate.
In our analysis leading to Figure~\ref{fig:dm} we have assumed that $\sigma_R$ is the lighter of the two components and is the dark matter. 

Notice that in this model the new gauge bosons can lead to signatures at the LHC, as well as novel flavor violating effects in the neutral meson systems 
such as K-, D- and B-mesons. Current data already imply stringent limits. A recent phenomenological study~\cite{Queiroz:2016gif} indicates that the masses 
of the new particles present in 3-3-1-1 models are expected to be high, so we take them heavy enough ($\gsim \mathcal{O}(10)$ TeV) so that they decouple at the electroweak scale.
In this case, apart from \sm particles, the only particles that are not decoupled at the electroweak scale are the two components of $\sigma$ which can be light, as they are \0331 singlets.
Under these simplifying approximations, the relic density of our dark matter candidate $\sigma_R$ is mainly controlled by its quartic coupling $\lambda_{\sigma h}$ with the \sm Higgs boson.
Furthermore, its ``effective interaction strength'' with the nuclei determining the dark matter--nucleus interaction cross section is also directly proportional 
to $\lambda_{\sigma h}$.

This simplified scenario has only one parameter, i.e. the $\lambda_{\sigma h}$ coupling, responsible for both the relic abundance as well as the direct detection cross section.
As a result, the constraints are rather tight, as can be seen from Figure \ref{fig:dm}.
This Figure indicates the restrictions on the $\lambda_{\sigma h} - m_{\sigma_R}$ plane obtained by requiring the correct dark matter relic abundance, as well as by imposing the dark matter direct detection constraints.
The mass of the dark matter $\sigma_R$ is confined to two distinct allowed regions. The first allowed region is near half the Higgs mass, where resonant 
annihilation of dark matter to the Higgs boson allows the relic density constraints to be satisfied for very small values of the coupling $\lambda_{\sigma h}$,
 well below the current direct detection bounds.
The second allowed region of $m_{\sigma_R}$ starts at around 1 TeV, where the direct detection constraints on the coupling $\lambda_{\sigma h}$ are weak. 
Thus, even within this constrained scenario, the field $\sigma_R$ can be a good dark matter candidate, provided its mass lies in one of these two allowed regions.

Before ending this section we wish to remark that Figure \ref{fig:dm} is plotted for a very constrained scenario with all but one coupling of the dark matter field set to zero.  This need not be the case. 
In the presence of other couplings, particularly the quartic coupling between         $\sigma_R$ and the other scalars, several additional channels for dark matter 
annihilation will open up. 
Thus the relic density constraints on the quartic coupling $\lambda_{\sigma h}$ can be substantially weakened,  opening up the allowed parameter space for $\lambda_{\sigma h}$ and $m_{\sigma_R}$. 
Thus, Figure \ref{fig:dm} should be taken as a kind of ``worst case scenario'' to illustrate consistency.
Finally, as we have stated before, $\sigma_R$ need not be the lightest $M_P$-odd particle in our model. 
A complete phenemenological study of all possible dark matter candidates is not the main aim of our paper and hence we will not explore in detail other possibilities.


\black
\section{Discussion}
\label{sec:discussion}


Many general phenomenological features involving the weak SU(3) gauge group, such as present in \tt1 schemes, are common also to our model. 
These theories imply the existence of new $Z^\prime$ gauge bosons that can be produced in proton-proton collisions through the Drell-Yan mechanism, leading to dilepton events at the LHC.
In addition, and more distinctively when compared to other electroweak extensions, the anomaly cancellation solution based on having one of the quark families transforming differently from the others implies the existence of flavor changing neutral currents at the tree level~\cite{Singer:1980sw}.
As a result one can have effects in the neutral meson systems such as K-, D- and B-mesons. Current LHC, Belle and BaBar data already imply stringent limits, discussed 
in~\cite{Queiroz:2016gif}.
Several other phenomenological aspects of this \3311 models were already discussed in Refs.~\cite{Alves:2016fqe} and~\cite{Dong:2014wsa}.

The main motivation of the current paper was the issue of dark matter. We note that in the present model all \sm fields have $M_P=1$, thus the lightest $M_P=-1$ is automatically stable and constitutes a potential WIMP dark matter candidate particle. Among the electrically neutral fields with $M_P=-1$, we have $\sigma$ and the lighter of the $F$'s, i.e. $(F_{aL,R})_1$.
Whichever is the lightest of these, can be a potential dark matter candidate. We discussed explicitly a benchmark for the scalar dark matter ($\sigma_R$) case.

As we noted, this model is characterized by the existence of extra fermions and scalar bosons needed for implementing the scotogenic scenario as well as for breaking the extended gauge symmetry. As a result one expects a plethora of possible collider signatures associated to the extra particles. Clearly, dedicated studies, similar to that in ~\cite{Diaz:2016udz}, would be required in order to scrutinize the associated detection potential at current and upcoming collider experiments, such as future runs of the LHC as well as future linear Colliders.

Finally, concerning the role of matter parity arising from the gauge sector in stabilizing the WIMP dark matter particle candidate, we note that this is a very general idea. Indeed, it may have alternative realizations from the one developed here. 


\acknowledgments
SKK and OP were supported by the National Research Foundation of Korea (NRF) grants 2009-0083526, 2017R1A2B4006338 and 2017K1A3A7A09016430. RS and JV are supported by the Spanish grants SEV-2014-0398 and FPA2017-85216-P (AEI/FEDER, UE), PROMETEO/2018/165 (Generalitat Valenciana) and the Spanish Red Consolider MultiDark FPA2017-90566-REDC. CAV-A is supported by the Mexican C\'atedras CONACYT project 749 and SNI 58928.
The relic abundance and direct detection constraints are calculated using the MicroOmegas package~\cite{Belanger:2018mqt}.

\appendix
\section{Anomaly Cancellation}
\label{sec:anomaly-cancellation}

``Right handed neutrinos'' $\nu_R$ with chiral charges were first discussed in context of $B-L$ symmetry in \cite{Montero:2007cd,Ma:2014qra,Ma:2015raa,Ma:2015mjd}. 
Here we show that, despite the non-trivial nature of the \3311 gauge symmetry characterizing our model, the unconventional U(1)$_N$ charges of $\nu_R$ $(-4,-4,5)$ ensure that the anomaly free conditions are fulfilled. \\

\textbf{Non-trivial anomalies}
\begin{itemize}
\item $[SU(3)_C]^2U(1)_X$:
\begin{equation}  
\begin{split}
\sum_{\mathrm{quarks}}& (X_{Q_L}-X_{Q_R})=2\times 3 X_{q_{\alpha L}}+3X_{q_{3L}}-3X_{u_{aR}}-3X_{d_{aR}}-X_{U_{3R}}-2X_{D_{\alpha R}}\\
&=6(0)+3(1/3)-3(2/3)-3(-1/3)-(2/3)-2(-1/3)=0,
\end{split}
\end{equation}

\item $[SU(3)_C]^2U(1)_N$:
\begin{equation}  
\begin{split} 
\sum_{\mathrm{quarks}} &(N_{Q_L}-N_{Q_R})=2\times 3 N_{q_{\alpha L}}+3N_{q_{3 L}}-3N_{u_{aR}}-3N_{d_{aR}}-N_{U_{3R}}-2N_{D_{\alpha R}}\\
&=6(0)+3(2/3)-3(1/3)-3(1/3)-(4/3)-2(-2/3)=0,
\end{split}
\end{equation}

\item $[SU(3)_L]^2U(1)_X$:
\begin{equation}  
\begin{split}
\sum_{\substack{ \mathrm{fermion} \\  \mathrm{(anti)triplets}}} &(X_{f_L}-X_{f_R})
=3X_{l_{aL}}+2\times 3 X_{q_{\alpha L}}+3X_{q_{3L}}+3 X_{F_{aL}}-3 X_{F_{aR}}\\
&=3(-1/3)+6(0)+3(1/3) +3(-1/3)-3 (-1/3)=0,
\end{split}
\end{equation}

\item $[SU(3)_L]^2U(1)_N$:
\begin{equation}  
\begin{split} 
\sum_{\substack{ \mathrm{fermion} \\  \mathrm{(anti)triplets}}} &(N_{f_L}-N_{f_R})
=3N_{l_{aL}}+2\times 3 N_{q_{\alpha L}}+3N_{q_{3L}}+3 N_{F_{aL}}-3 N_{F_{aR}}\\
&=3(-2/3)+6(0)+3(2/3)+3(-1/3)-3 (-1/3)=0,
\end{split}
\end{equation}

\item $[U(1)_X]^2U(1)_N$:
\begin{equation}  
\begin{split} 
\sum_{\mathrm{fermions}}&(X^2_{f_L}N_{f_L}-X^2_{f_R}N_{f_R})
=3\times 3 X^2_{l_{aL}}N_{l_{aL}}+2\times 3\times 3 X^2_{q_{\alpha L}}N_{q_{\alpha L}}+3\times 3 X^2_{q_{3L}} N_{q_{3L}}\\
&-3\times 3 X^2_{u_{aR}}N_{u_{aR}}-3\times 3 X^2_{d_{aR}}N_{d_{aR}}-3X^2_{U_{3R}} N_{U_{3R}}-2\times 3 X^2_{D_{\alpha R}}N_{D_{\alpha R}}\\
&-3X^2_{e_{aR}} N_{e_{aR}}-2X^2_{\nu_{\alpha R}} N_{\nu_{\alpha R}}-X^2_{\nu_{3R}} N_{\nu_{3R}}+3 X^2_{F_{aL}}N_{F_{aL}}-3 X^2_{F_{aR}}N_{F_{aR}}\\
=&9(-1/3)^2(-2/3)+18 (0)^2(0)+9 (1/3)^2(2/3)-9 (2/3)^2(1/3)\\
&-9 (-1/3)^2(1/3)-3(2/3)^2(4/3)-6 (-1/3)^2(-2/3)\\
&-3(-1)^2(-1)-2 (0)^2 (-4) -  (0)^2 (5)\\
&+3(-1/3)^2(-1/3)-3(-1/3)^2(-1/3)=0,
\end{split}
\end{equation}

\item $U(1)_X[U(1)_N]^2$:
\begin{equation}  
\begin{split} 
\sum_{\mathrm{fermions}}&(X_{f_L}N^2_{f_L}-X_{f_R}N^2_{f_R})
=3\times 3 X_{l_{aL}}N^2_{l_{aL}}+2\times 3\times 3 X_{q_{\alpha L}}N^2_{q_{\alpha L}}+3\times 3 X_{q_{3L}} N^2_{q_{3L}}\\
&-3\times 3 X_{u_{aR}}N^2_{u_{aR}}-3\times 3 X_{d_{aR}}N^2_{d_{aR}}-3X_{U_{3R}} N^2_{U_{3R}}-2\times 3 X_{D_{\alpha R}}N^2_{D_{\alpha R}}\\
&-3X_{e_{aR}} N^2_{e_{aR}}-2X_{\nu_{\alpha R}} N^2_{\nu_{\alpha R}}-X_{\nu_{3R}} N^2_{\nu_{3R}}+3 X_{F_{aL}}N^2_{F_{aL}}-3 X_{F_{aR}}N^2_{F_{aR}}\\
=&9(-1/3)(-2/3)^2+18 (0)(0)^2+9 (1/3)(2/3)^2-9 (2/3)(1/3)^2\\
&-9 (-1/3)(1/3)^2-3(2/3)(4/3)^2-6 (-1/3)(-2/3)^2\\
&-3(-1)(-1)^2 -2(0)(-4)^2- (0)(5)^2\\
&+3(-1/3)(-1/3)^2-3(-1/3)(-1/3)^2=0,
\end{split}
\end{equation}

\item $[U(1)_X]^3$:
\begin{equation}  
\begin{split} 
\sum_{\mathrm{fermions}}&(X_{f_L}^3-X_{f_R}^3)
=3\times 3 X^3_{l_{aL}}+2\times 3\times 3 X^3_{q_{\alpha L}}+3\times 3 X^3_{q_{3L}} \\
&-3\times 3 X^3_{u_{aR}}-3\times 3 X_{d_{aR}}^3-3X_{U_{3R}} ^3-2\times 3 X_{D_{\alpha R}}^3\\
&-3X_{e_{aR}} ^3-2X_{\nu_{\alpha R}}^3-X_{\nu_{3R}}^3+3 X_{F_{aL}}^3-3 X_{F_{aR}}^3\\
=&9(-1/3)^3+18 (0)^3+9 (1/3)^3-9 (2/3)^3\\
&-9 (-1/3)^3-3(2/3)^3-6 (-1/3)^3\\
&-3(-1)^3 -2(0)^3- (0)^3\\
&+3(-1/3)^3-3(-1/3)^3=0,
\end{split}
\end{equation}

\item $[U(1)_N]^3$
\begin{equation}  
\begin{split} 
\sum_{\mathrm{fermions}}&(N^3_{f_L}-N^3_{f_R})
=3\times 3 N^3_{l_{aL}}+2\times 3\times 3N^3_{q_{\alpha L}}+3\times 3 N^3_{q_{3L}}\\
&-3\times 3 N^3_{u_{aR}}-3\times 3 N^3_{d_{aR}}-3N^3_{U_{3R}}-2\times 3 N^3_{D_{\alpha R}}\\
&-3 N^3_{e_{aR}}-2N^3_{\nu_{\alpha R}}-N^3_{\nu_{3R}}+3 N^3_{F_{aL}}-3 N^3_{F_{aR}}\\
=&9(-2/3)^3+18 (0)^3+9(2/3)^3-9 (1/3)^3\\
&-9(1/3)^3-3(4/3)^3-6 (-2/3)^3\\
&-3(-1)^3 -2(-4)^3- (5)^3\\
&+3(-1/3)^3-3(-1/3)^3=0,
\end{split}
\end{equation}

\item $[\mathrm{Grav}]U(1)_X$:
\begin{equation}  
\begin{split} 
\sum_{\mathrm{fermions}}&(X_{f_L}-X_{f_R})
=3\times 3 X_{l_{aL}}+2\times 3\times 3 X_{q_{\alpha L}}+3\times 3 X_{q_{3L}} \\
&-3\times 3 X_{u_{aR}}-3\times 3 X_{d_{aR}}-3X_{U_{3R}}-2\times 3 X_{D_{\alpha R}}\\
&-3X_{e_{aR}}-2X_{\nu_{\alpha R}}-X_{\nu_{3R}}+3 X_{F_{aL}}-3 X_{F_{aR}}\\
=&9(-1/3)+18 (0)+9 (1/3)-9 (2/3)\\
&-9 (-1/3)-3(2/3)-6 (-1/3)\\
&-3(-1) -2(0)- (0)\\
&+3(-1/3)-3(-1/3)=0,
\end{split}
\end{equation}

\item $[\mathrm{Grav}]U(1)_N$
\begin{equation}  
\begin{split} 
\sum_{\mathrm{fermions}}&(N_{f_L}-N_{f_R})
=3\times 3 N_{l_{aL}}+2\times 3\times 3N_{q_{\alpha L}}+3\times 3 N_{q_{3L}}\\
&-3\times 3 N_{u_{aR}}-3\times 3 N_{d_{aR}}-3N_{U_{3R}}-2\times 3 N_{D_{\alpha R}}\\
&-3 N_{e_{aR}}-2N_{\nu_{\alpha R}}-N_{\nu_{3R}}+3 N_{F_{aL}}-3 N_{F_{aR}}\\
=&9(-2/3)+18 (0)+9(2/3)-9 (1/3)\\
&-9(1/3)-3(4/3)-6 (-2/3)\\
&-3(-1) -2(-4)- (5)\\
&+3(-1/3)-3(-1/3)=0.
\end{split}
\end{equation}

\end{itemize}

\section{Scalar mass spectrum}
\label{sec:mass_det}

\begin{enumerate}
 
\item[ I.] $Q=0,M_P=+$, CP = even: \\
The mass squared matrix elements $m^2_{ij}$; $i,j = 1 \cdots 7$ in the basis 
$(\eta_{1R},\rho_{2R},\chi_{3R},\phi_R,\Omega_{11R},\Omega_{33R},S_R)$ are given as
\begin{align}
\label{eq:m20pR}
&2 m^2_{11}= 2\lambda_{\eta}v_1^2-\sqrt{2}\mu_1 \frac{v_2 w}{v_1} 
&2 m^2_{12}= 2\lambda_{\eta\rho}v_1 v_2 + \sqrt{2}\mu_1 w \\
&2 m^2_{13}= 2\lambda_{\eta\chi}v_1 w + \sqrt{2}\mu_1 v_2 
&2 m^2_{14}= 2\lambda_{\eta\phi}v_1\Lambda \nonumber \\
&2 m^2_{15}= 2\left(\lambda_{\eta\Omega}+\lambda_{\eta\Omega 2}\right)w_1 v_1 
&2 m^2_{16}= 2\lambda_{\eta\Omega}w_2 v_1 \nonumber \\
&2 m^2_{17}= 2\lambda_{\eta s}v_s v_1 
&2 m^2_{22}= 2\lambda_{\rho}v_2^2-\sqrt{2}\mu_1 \frac{v_1 w}{v_2} \nonumber \\
&2 m^2_{23}= 2\lambda_{\rho\chi}v_2 w+\sqrt{2}\mu_1 v_1 
&2 m^2_{24}= 2\lambda_{\rho\phi}v_2\Lambda \nonumber \\
&2 m^2_{25}= 2\lambda_{\rho\Omega}w_1 v_2 
&2 m^2_{26}= 2\lambda_{\rho\Omega}w_2 v_2 \nonumber \\
&2 m^2_{27}= 2\lambda_{\rho s}v_s v_2 
&2 m^2_{33}= 2\lambda_{\chi}w^2-\sqrt{2}\mu_1 \frac{v_1 v_2}{w} \nonumber \\
&2 m^2_{34}= 2\lambda_{\chi\phi} w\Lambda 
&2 m^2_{35}= 2\lambda_{\chi\Omega}w_1 w \nonumber \\
&2 m^2_{36}= 2\left(\lambda_{\chi\Omega}+\lambda_{\chi\Omega 2}\right)w_2 w +2\lambda_1 v_s w 
&2 m^2_{37}= 2\lambda_{\chi s}v_s w +2\lambda_1 w_2 w \nonumber \\
&2 m^2_{44}= 2\lambda_{\phi} \Lambda^2-\lambda_3 \frac{v_s^3}{\Lambda} 
&2 m^2_{45}= 2\lambda_{\phi\Omega}w_1\Lambda \nonumber \\
&2 m^2_{46}= 2\lambda_{\phi\Omega}w_2\Lambda 
&2 m^2_{47}= 2\lambda_{\phi s}v_s\Lambda+3\lambda_3 v_s^2 \nonumber \\
&2 m^2_{55}= 2\lambda_{\Omega}w_1^2 
&2 m^2_{56}= 2\lambda_{\Omega}w_1 w_2 +2\frac{\lambda_{\eta\Omega 2}w_1 w_2 v_1^2-\left(\lambda_{\chi\Omega 2} w_2 +\lambda_1 v_s\right)w_1w^2}{w_1^2-w_2^2} \nonumber \\
&2 m^2_{57}= 2\lambda_{s\Omega} v_s w_1 
&2 m^2_{66}= 2\lambda_{\Omega}w_2^2-\lambda_1\frac{v_s w^2}{w_2} \nonumber \\
&2 m^2_{67}= 2\lambda_{s\Omega}v_s w_2+\lambda_1 w^2 
&2 m^2_{77}= 2\lambda_s v_s^2+3\lambda_3 v_s \Lambda-\lambda_1 \frac{w_2 w^2}{v_s} \nonumber 
\end{align}

\item[ II.] $Q=0,M_P=+$, CP = odd: \\
The mass squared matrix in the basis $(\eta_{1I},\rho_{2I},\chi_{3I},\phi_I,\Omega_{33I},S_I)$ is given as 
\begin{equation}
\label{eq:m20pI}
\frac{1}{2}\left(\begin{matrix}
-\sqrt{2}\mu_1 \frac{v_2 w}{v_1} & -\sqrt{2}\mu_1 w & -\sqrt{2}\mu_1 v_2 & 0 & 0 & 0 \\
-\sqrt{2}\mu_1 w & -\sqrt{2}\mu_1 \frac{v_1 w}{v_2} & -\sqrt{2}\mu_1 v_1 & 0 & 0 & 0 \\
-\sqrt{2}\mu_1 v_2 & -\sqrt{2}\mu_1 v_1 & -\sqrt{2}\mu_1 \frac{v_1 v_2}{w}-4\lambda_1 w_2 v_s & 0 & -2 \lambda_1 v_s w & -2 \lambda_1 w_2 w \\ 
0 & 0 & 0 & -\lambda_3 \frac{v_s^3}{\Lambda} & 0 & 3\lambda_3 v_s^2 \\
0 & 0 & -2 \lambda_1 v_s w & 0 & -\lambda_1 \frac{v_s w^2}{w_2} & -\lambda_1 w^2 \\
0 & 0 & -2 \lambda_1 w_2 w & 3\lambda_3 v_s^2 & -\lambda_1 w^2 & -\lambda_1 \frac{w_2 w^2}{v_s}
\end{matrix}\right).
\end{equation}

\item[ III.] $Q=0,M_P=-$, CP = even: \\
The mass squared matrix in the basis $(\eta_{3R}, \sigma_R, \chi_{1R}, \Omega_{13R})$ is given as 
\begin{align}
\label{eq:m20mR}
\scriptstyle{
\frac{1}{2}\left(\begin{matrix}
-\lambda_{\eta\Omega 2}\left(w_1^2-w_2^2\right)+\lambda_{\eta\chi 2}w^2-\sqrt{2}\mu_1 \frac{v_2 w}{v_1} & \left(\lambda_2 w_2+\lambda_4 v_s\right)w & \lambda_{\eta\chi 2}v_1 w-\sqrt{2}\mu_1v_2 & \lambda_{\eta\Omega 2} \left(w_1+w_2\right)v_1 \\
\left(\lambda_2 w_2+\lambda_4 v_s\right)w & m^2_{22} & \left(\lambda_{2}w_1+\lambda_4 v_s\right)v_1 & \lambda_2 v_1 w \\
\lambda_{\eta\chi 2}v_1 w-\sqrt{2}\mu_1v_2 & \left(\lambda_{2}w_1+\lambda_4 v_s\right)v_1 & m^2_{33} & \left(\lambda_{\chi\Omega 2}\left(w_1+w_2\right)+2 \lambda_1 v_s\right)w \\
\lambda_{\eta\Omega 2} \left(w_1+w_2\right)v_1 & \lambda_2 v_1 w & \left(\lambda_{\chi\Omega 2}\left(w_1+w_2\right)+2 \lambda_1 v_s\right)w & m^2_{44}
\end{matrix}\right),
}
\end{align}
where
\begin{align*}
m^2_{22}&=2 m^2_{\sigma}+\lambda_{\sigma\Omega}\left(w_1^2+w_2^2\right)+\lambda_{s\sigma}v_s^2+\lambda_{\eta\sigma}v_1^2+\lambda_{\rho\sigma}v_2^2+\lambda_{\phi\sigma}\Lambda^2+\lambda_{\chi\sigma}w^2+2\sqrt{2}\mu_2 v_s, \\
m^2_{33}&=\lambda_{\chi\Omega 2}\left(w_1^2-w_2^2\right)+2 \lambda_1 \left(w_1-w_2\right)v_s +\lambda_{\eta\chi 2}v_1^2-\sqrt{2}\mu_1 \frac{v_1 v_2}{w}, \\
m^2_{44}&=\frac{\left(\lambda_{\chi\Omega 2}\left(w_1+w_2\right)+2\lambda_1 v_s\right)w^2-\lambda_{\eta\Omega 2}\left(w_1+w_2\right)v_1^2}{w_1-w_2},
\end{align*}

\item[ IV.] $Q=0,M_P=-$, CP = odd: \\
The mass squared matrix in the basis $(\eta_{3I}, \sigma_I, \chi_{1I}, \Omega_{13I})$ is given as
\begin{align}
\label{eq:m20mI}
\scriptstyle{
\frac{1}{2}\left(\begin{matrix}
\lambda_{\eta\chi 2}w^2-\lambda_{\eta\Omega 2}\left(w_1^2-w_2^2\right)-\sqrt{2}\mu_1 \frac{v_2 w}{v_1} & \left(\lambda_{2}w_2+\lambda_4 v_s\right)w & -\lambda_{\eta\chi 2}v_1 w +\sqrt{2}\mu_1 v_2 & \lambda_{\eta\Omega 2}\left(-w_1+w_2\right)v_1 \\
\left(\lambda_{2}w_2+\lambda_4 v_s\right)w & m^2_{22} & \left(\lambda_{2}w_1-\lambda_4 v_s\right)v_1 & \lambda_2 v_1 w \\
-\lambda_{\eta\chi 2}v_1 w +\sqrt{2}\mu_1 v_2 & \left(\lambda_{2}w_1-\lambda_4 v_s\right)v_1 & m^2_{33} & \lambda_{\chi\Omega 2}\left(w_1-w_2\right)w -2\lambda_1 v_s w \\
\lambda_{\eta\Omega 2}\left(-w_1+w_2\right)v_1 & \lambda_2 v_1 w & \lambda_{\chi\Omega 2}\left(w_1-w_2\right)w -2\lambda_1 v_s w & m^2_{44}
\end{matrix}\right),
}
\end{align}
where 
\begin{align*}
m^2_{22}&=2 m_{\sigma}^2+\lambda_{\sigma\Omega}\left(w_1^2+w_2^2\right)+\lambda_{s\sigma}v_s^2+\lambda_{\eta\sigma}v_1^2+\lambda_{\rho\sigma}v_2^2+\lambda_{\phi\sigma}\Lambda^2+\lambda_{\chi\sigma}w^2-2\sqrt{2}\mu_2 v_s, \\
m^2_{33}&= \lambda_{\chi\Omega 2}\left(w_1^2-w_2^2\right)-2\lambda_1\left(w_1+w_2\right)v_s+\lambda_{\eta\chi 2}v_1^2-\sqrt{2}\mu_1\frac{v_1 v_2}{w},\\
m^2_{44}&= \frac{\left(\lambda_{\chi\Omega 2}\left(w_1-w_2\right)-2\lambda_1 v_s\right)w^2-\lambda_{\eta\Omega 2}\left(w_1-w_2\right)v_1^2}{w_1+w_2},
\end{align*}

\end{enumerate}

\bibliographystyle{utphys}
\bibliography{bibliography} 

\providecommand{\href}[2]{#2}\begingroup\raggedright\begin{thebibliography}{10}

\bibitem{Jungman:1995df}
G.~Jungman, M.~Kamionkowski, and K.~Griest, ``{Supersymmetric dark matter},''
  {\em Phys.Rept.} {\bfseries 267} 195--373,
  \href{http://arxiv.org/abs/hep-ph/9506380}{{\ttfamily arXiv:hep-ph/9506380
  [hep-ph]}}.

\bibitem{Boucenna:2014zba}
S.~M. Boucenna, S.~Morisi, and J.~W.~F. Valle, ``{The low-scale approach to
  neutrino masses},'' \href{http://dx.doi.org/10.1155/2014/831598}{{\em Adv.
  High Energy Phys.} {\bfseries 2014} (2014) 831598},
  \href{http://arxiv.org/abs/1404.3751}{{\ttfamily arXiv:1404.3751 [hep-ph]}}.

\bibitem{Ma:2006km}
E.~Ma, ``{Verifiable radiative seesaw mechanism of neutrino mass and dark
  matter},'' {\em Phys.Rev.} {\bfseries D73} 077301,
  \href{http://arxiv.org/abs/hep-ph/0601225}{{\ttfamily arXiv:hep-ph/0601225
  [hep-ph]}}.

\bibitem{Hirsch:2013ola}
M.~Hirsch {\em et~al.}, ``{WIMP dark matter as radiative neutrino mass
  messenger},'' {\em JHEP} {\bfseries 1310} 149,
  \href{http://arxiv.org/abs/1307.8134}{{\ttfamily arXiv:1307.8134 [hep-ph]}}.

\bibitem{Merle:2016scw}
A.~Merle {\em et~al.}, ``{Consistency of WIMP Dark Matter as radiative neutrino
  mass messenger},'' \href{http://dx.doi.org/10.1007/JHEP07(2016)013}{{\em
  JHEP} {\bfseries 07} (2016) 013},
  \href{http://arxiv.org/abs/1603.05685}{{\ttfamily arXiv:1603.05685
  [hep-ph]}}.

\bibitem{Alves:2016fqe}
A.~Alves {\em et~al.}, ``{Matter-parity as a residual gauge symmetry: Probing a
  theory of cosmological dark matter},''
  \href{http://dx.doi.org/10.1016/j.physletb.2017.07.056}{{\em Phys. Lett.}
  {\bfseries B772} (2017) 825--831},
\href{http://arxiv.org/abs/1612.04383}{{\ttfamily arXiv:1612.04383 [hep-ph]}}.

\bibitem{Dong:2017zxo}
P.~V. Dong {\em et~al.}, ``{The Dark Side of Flipped Trinification},''
  \href{http://dx.doi.org/10.1007/JHEP04(2018)143}{{\em JHEP} {\bfseries 04}
  (2018) 143},
\href{http://arxiv.org/abs/1710.06951}{{\ttfamily arXiv:1710.06951 [hep-ph]}}.

\bibitem{Singer:1980sw}
M.~Singer, J.~W.~F. Valle, and J.~Schechter, ``{Canonical Neutral Current
  Predictions From the Weak Electromagnetic Gauge Group SU(3) X $u$(1)},''
  \href{http://dx.doi.org/10.1103/PhysRevD.22.738}{{\em Phys.Rev.} {\bfseries
  D22} (1980) 738}.

\bibitem{Pisano:1991ee}
F.~Pisano and V.~Pleitez, ``{An SU(3) x U(1) model for electroweak
  interactions},'' \href{http://dx.doi.org/10.1103/PhysRevD.46.410}{{\em Phys.
  Rev.} {\bfseries D46} (1992) 410--417},
\href{http://arxiv.org/abs/hep-ph/9206242}{{\ttfamily arXiv:hep-ph/9206242
  [hep-ph]}}.

\bibitem{Frampton:1992wt}
P.~H. Frampton, ``{Chiral dilepton model and the flavor question},''
\href{http://dx.doi.org/10.1103/PhysRevLett.69.2889}{{\em Phys. Rev. Lett.}
  {\bfseries 69} (1992) 2889--2891}.

\bibitem{Hernandez:2013mcf}
A.~E. Carcamo~Hernandez, R.~Martinez, and F.~Ochoa, ``{Radiative seesaw-type
  mechanism of quark masses in $SU(3)_C \otimes SU(3)_L \otimes U(1)_X$},''
  \href{http://dx.doi.org/10.1103/PhysRevD.87.075009}{{\em Phys. Rev.}
  {\bfseries D87} no.~7, (2013) 075009},
\href{http://arxiv.org/abs/1302.1757}{{\ttfamily arXiv:1302.1757 [hep-ph]}}.

\bibitem{Hernandez:2014lpa}
A.~E. Carcamo~Hernandez, E.~Catano~Mur, and R.~Martinez, ``{Lepton masses and
  mixing in $SU(3)_{C}\otimes SU(3)_{L}\otimes U(1)_{X}$ models with a $S_3$
  flavor symmetry},'' \href{http://dx.doi.org/10.1103/PhysRevD.90.073001}{{\em
  Phys. Rev.} {\bfseries D90} no.~7, (2014) 073001},
\href{http://arxiv.org/abs/1407.5217}{{\ttfamily arXiv:1407.5217 [hep-ph]}}.

\bibitem{CarcamoHernandez:2017cwi}
A.~E. Carcamo~Hernandez, S.~Kovalenko, H.~N. Long, and I.~Schmidt, ``{A variant
  of 3-3-1 model for the generation of the SM fermion mass and mixing
  pattern},'' \href{http://dx.doi.org/10.1007/JHEP07(2018)144}{{\em JHEP}
  {\bfseries 07} (2018) 144},
\href{http://arxiv.org/abs/1705.09169}{{\ttfamily arXiv:1705.09169 [hep-ph]}}.

\bibitem{Barreto:2017xix}
E.~R. Barreto, A.~G. Dias, J.~Leite, C.~C. Nishi, R.~L.~N. Oliveira, and W.~C.
  Vieira, ``{Hierarchical fermions and detectable Z' from effective
  two-Higgs-triplet 3-3-1 model},''
  \href{http://dx.doi.org/10.1103/PhysRevD.97.055047}{{\em Phys. Rev.}
  {\bfseries D97} no.~5, (2018) 055047},
\href{http://arxiv.org/abs/1709.09946}{{\ttfamily arXiv:1709.09946 [hep-ph]}}.

\bibitem{Deppisch:2016jzl}
F.~F. Deppisch {\em et~al.}, ``{331 Models and Grand Unification: From Minimal
  SU(5) to Minimal SU(6)},''
  \href{http://dx.doi.org/10.1016/j.physletb.2016.10.002}{{\em Phys. Lett.}
  {\bfseries B762} (2016) 432--440},
  \href{http://arxiv.org/abs/1608.05334}{{\ttfamily arXiv:1608.05334
  [hep-ph]}}.

\bibitem{Hati:2017aez}
C.~Hati {\em et~al.}, ``{Towards gauge coupling unification in left-right
  symmetric $\mathrm{SU(3)_c \times SU(3)_L \times SU(3)_R \times U(1)_{X}}$
  theories},'' \href{http://arxiv.org/abs/1703.09647}{{\ttfamily
  arXiv:1703.09647 [hep-ph]}}.

\bibitem{Dong:2014wsa}
P.~V. Dong, D.~T. Huong, F.~S. Queiroz, and N.~T. Thuy, ``{Phenomenology of the
  3-3-1-1 model},'' \href{http://dx.doi.org/10.1103/PhysRevD.90.075021}{{\em
  Phys. Rev.} {\bfseries D90} no.~7, (2014) 075021},
\href{http://arxiv.org/abs/1405.2591}{{\ttfamily arXiv:1405.2591 [hep-ph]}}.

\bibitem{Dong:2015yra}
P.~V. Dong, ``{Unifying the electroweak and B-L interactions},''
  \href{http://dx.doi.org/10.1103/PhysRevD.92.055026}{{\em Phys. Rev.}
  {\bfseries D92} no.~5, (2015) 055026},
\href{http://arxiv.org/abs/1505.06469}{{\ttfamily arXiv:1505.06469 [hep-ph]}}.

\bibitem{Bonilla:2018ynb}
C.~Bonilla, S.~Centelles-Chulia, R.~Cepedello, E.~Peinado, and R.~Srivastava,
  ``{Dark matter stability and Dirac neutrinos using only Standard Model
  symmetries},''
\href{http://arxiv.org/abs/1812.01599}{{\ttfamily arXiv:1812.01599 [hep-ph]}}.

\bibitem{CentellesChulia:2019gic}
S.~Centelles~Chulia, R.~Cepedello, E.~Peinado, and R.~Srivastava, ``{Scotogenic
  Dark Symmetry as a residual subgroup of Standard Model Symmetries},''
\href{http://arxiv.org/abs/1901.06402}{{\ttfamily arXiv:1901.06402 [hep-ph]}}.

\bibitem{Ma:2019yfo}
E.~Ma, ``{Scotogenic $U(1)_\chi$ Dirac Neutrinos},''
\href{http://arxiv.org/abs/1901.09091}{{\ttfamily arXiv:1901.09091 [hep-ph]}}.

\bibitem{Ma:2015xla}
E.~Ma, ``{Derivation of Dark Matter Parity from Lepton Parity},''
  \href{http://dx.doi.org/10.1103/PhysRevLett.115.011801}{{\em Phys. Rev.
  Lett.} {\bfseries 115} no.~1, (2015) 011801},
\href{http://arxiv.org/abs/1502.02200}{{\ttfamily arXiv:1502.02200 [hep-ph]}}.

\bibitem{Montero:2007cd}
J.~C. Montero and V.~Pleitez, ``{Gauging U(1) symmetries and the number of
  right-handed neutrinos},''
  \href{http://dx.doi.org/10.1016/j.physletb.2009.03.065}{{\em Phys. Lett.}
  {\bfseries B675} (2009) 64--68},
\href{http://arxiv.org/abs/0706.0473}{{\ttfamily arXiv:0706.0473 [hep-ph]}}.

\bibitem{Ma:2014qra}
E.~Ma and R.~Srivastava, ``{Dirac or inverse seesaw neutrino masses with $B-L$
  gauge symmetry and $S_3$ flavor symmetry},''
  \href{http://dx.doi.org/10.1016/j.physletb.2014.12.049}{{\em Phys. Lett.}
  {\bfseries B741} (2015) 217--222},
\href{http://arxiv.org/abs/1411.5042}{{\ttfamily arXiv:1411.5042 [hep-ph]}}.

\bibitem{Ma:2015raa}
E.~Ma and R.~Srivastava, ``{Dirac or inverse seesaw neutrino masses from gauged
  $B–L$ symmetry},'' \href{http://dx.doi.org/10.1142/S0217732315300207}{{\em
  Mod. Phys. Lett.} {\bfseries A30} no.~26, (2015) 1530020},
\href{http://arxiv.org/abs/1504.00111}{{\ttfamily arXiv:1504.00111 [hep-ph]}}.

\bibitem{Ma:2015mjd}
E.~Ma, N.~Pollard, R.~Srivastava, and M.~Zakeri, ``{Gauge $B-L$ Model with
  Residual $Z_3$ Symmetry},''
  \href{http://dx.doi.org/10.1016/j.physletb.2015.09.010}{{\em Phys. Lett.}
  {\bfseries B750} (2015) 135--138},
\href{http://arxiv.org/abs/1507.03943}{{\ttfamily arXiv:1507.03943 [hep-ph]}}.

\bibitem{Valle:1983dk}
J.~W.~F. Valle and M.~Singer, ``{Lepton Number Violation With Quasi Dirac
  Neutrinos},'' \href{http://dx.doi.org/10.1103/PhysRevD.28.540}{{\em
  Phys.Rev.} {\bfseries D28} (1983) 540}.

\bibitem{Reig:2016ewy}
M.~Reig, J.~W.~F. Valle, and C.~A. Vaquera-Araujo, ``{Realistic
  $\mathrm{SU(3)_c \otimes SU(3)_L \otimes U(1)_X}$ model with a type II Dirac
  neutrino seesaw mechanism},''
  \href{http://dx.doi.org/10.1103/PhysRevD.94.033012}{{\em Phys. Rev.}
  {\bfseries D94} no.~3, (2016) 033012},
  \href{http://arxiv.org/abs/1606.08499}{{\ttfamily arXiv:1606.08499
  [hep-ph]}}.

\bibitem{GonzalezGarcia:1989py}
M.~Gonzalez-Garcia and J.~W.~F. Valle, ``{Cosmological Constraints on
  Additional Light Neutrinos and Neutral Gauge Bosons},''
  \href{http://dx.doi.org/10.1016/0370-2693(90)90426-7}{{\em Phys.Lett.}
  {\bfseries B240} (1990) 163}.

\bibitem{Dong:2013wca}
P.~V. Dong, H.~T. Hung, and T.~D. Tham, ``{3-3-1-1 model for dark matter},''
  \href{http://dx.doi.org/10.1103/PhysRevD.87.115003}{{\em Phys. Rev.}
  {\bfseries D87} no.~11, (2013) 115003},
\href{http://arxiv.org/abs/1305.0369}{{\ttfamily arXiv:1305.0369 [hep-ph]}}.

\bibitem{Aprile:2018dbl}
{\bfseries XENON} Collaboration, E.~Aprile {\em et~al.}, ``{Dark Matter Search
  Results from a One Ton-Year Exposure of XENON1T},''
  \href{http://dx.doi.org/10.1103/PhysRevLett.121.111302}{{\em Phys. Rev.
  Lett.} {\bfseries 121} no.~11, (2018) 111302},
\href{http://arxiv.org/abs/1805.12562}{{\ttfamily arXiv:1805.12562
  [astro-ph.CO]}}.

\bibitem{Akerib:2016vxi}
{\bfseries LUX} Collaboration, D.~S. Akerib {\em et~al.}, ``{Results from a
  search for dark matter in the complete LUX exposure},''
  \href{http://dx.doi.org/10.1103/PhysRevLett.118.021303}{{\em Phys. Rev.
  Lett.} {\bfseries 118} no.~2, (2017) 021303},
\href{http://arxiv.org/abs/1608.07648}{{\ttfamily arXiv:1608.07648
  [astro-ph.CO]}}.

\bibitem{Aghanim:2018eyx}
{\bfseries Planck} Collaboration, N.~Aghanim {\em et~al.}, ``{Planck 2018
  results. VI. Cosmological parameters},''
\href{http://arxiv.org/abs/1807.06209}{{\ttfamily arXiv:1807.06209
  [astro-ph.CO]}}.

\bibitem{Queiroz:2016gif}
F.~S. Queiroz, C.~Siqueira, and J.~W.~F. Valle, ``{Constraining Flavor Changing
  Interactions from LHC Run-2 Dilepton Bounds with Vector Mediators},''
  \href{http://dx.doi.org/10.1016/j.physletb.2016.10.057}{{\em Phys. Lett.}
  {\bfseries B763} (2016) 269--274},
\href{http://arxiv.org/abs/1608.07295}{{\ttfamily arXiv:1608.07295 [hep-ph]}}.

\bibitem{Diaz:2016udz}
M.~A. Diaz {\em et~al.}, ``{Heavy Higgs Boson Production at Colliders in the
  Singlet-Triplet Scotogenic Dark Matter Model},''
\href{http://arxiv.org/abs/1612.06569}{{\ttfamily arXiv:1612.06569 [hep-ph]}}.

\bibitem{Belanger:2018mqt}
G.~Bélanger, F.~Boudjema, A.~Goudelis, A.~Pukhov, and B.~Zaldivar,
  ``{micrOMEGAs5.0 : Freeze-in},''
  \href{http://dx.doi.org/10.1016/j.cpc.2018.04.027}{{\em Comput. Phys.
  Commun.} {\bfseries 231} (2018) 173--186},
\href{http://arxiv.org/abs/1801.03509}{{\ttfamily arXiv:1801.03509 [hep-ph]}}.

\end{thebibliography}\endgroup
\end{document}